\documentclass[11pt,twoside]{article}
\usepackage{./asp2014}

\aspSuppressVolSlug
\resetcounters

\begin{document}

\title{Resolving the Structure and Kinematics of the Youngest HII Regions and Radio Jets from Young Stellar Objects}
\author{Roberto Galv\'an-Madrid$^1$, Maite Beltr\'an$^2$, Adam Ginsburg$^3$, Carlos Carrasco-Gonz\'alez$^1$, Hauyu Baobab Liu$^4$, Luis F. Rodr\'iguez$^1$, Stanley Kurtz$^1$
\affil{$^1$Instituto de Radioastronom\'ia y Astrof\'isica,
            Universidad Nacional Aut\'onoma de M\'exico,
            Apdo. Postal 72-3 (Xangari), Morelia, 
            Michoac\'an 58089, Mexico.; \email{r.galvan@irya.unam.mx}}
\affil{$^2$INAF-Osservatorio Astrofisico di Arcetri, Largo E. Fermi 5, I-50125 Firenze, Italy.}
\affil{$^3$National Radio Astronomy Observatory, 1003 Lopezville Rd., Socorro, NM 87801, USA.}
\affil{$^4$European Southern Observatory, Karl-Schwarzschild-Stra\ss e 2, D-85748 Garching bei M\"unchen, Germany.}}

\paperauthor{Roberto Galv\'an-Madrid}{r.galvan@irya.unam.mx}{0000-0003-1480-4643}{Universidad Nacional Aut\'onoma de M\'exico}{Instituto de Radioastronom\'ia y Astrof\'isica}{Morelia}{Michoac\'an}{58089}{Mexico}
\paperauthor{Maite Beltr\'an}{mbeltran@arcetri.astro.it}{0000-0003-3315-5626}{INAF}{Osservatorio Astrofisico di Arcetri}{Firenze}{}{I-50125}{Italy}
\paperauthor{Adam Ginsburg}{adam.g.ginsburg@gmail.com}{0000-0001-6431-9633}{National Radio Astronomy Observatory}{Array Operations Center}{Socorro}{New Mexico}{87801}{USA}
\paperauthor{Carlos Carrasco-Gonz\'alez}{c.carrasco@irya.unam.mx}{}{Universidad Nacional Aut\'onoma de M\'exico}{Instituto de Radioastronom\'ia y Astrof\'isica}{Morelia}{Michoac\'an}{58089}{Mexico}
\paperauthor{Hauyu Baobab Liu}{baobabyoo@gmail.com}{0000-0003-2300-2626}{European Southern Observatory}{Headquarters}{Garching}{Bavaria}{85748}{Germany}
\paperauthor{Luis F. Rodr\'iguez}{l.rodriguez@irya.unam.mx}{}{Universidad Nacional Aut\'onoma de M\'exico}{Instituto de Radioastronom\'ia y Astrof\'isica}{Morelia}{Michoac\'an}{58089}{Mexico}
\paperauthor{Stanley Kurtz}{s.kurtz@irya.unam.mx}{}{Universidad Nacional Aut\'onoma de M\'exico}{Instituto de Radioastronom\'ia y Astrof\'isica}{Morelia}{Michoac\'an}{58089}{Mexico}


\section{Introduction}

\subsection{Free-free and Recombination Line Emission from High-Mass Young Stellar Objects}

\begin{figure*}
\begin{center}
\includegraphics[width=0.9\textwidth]{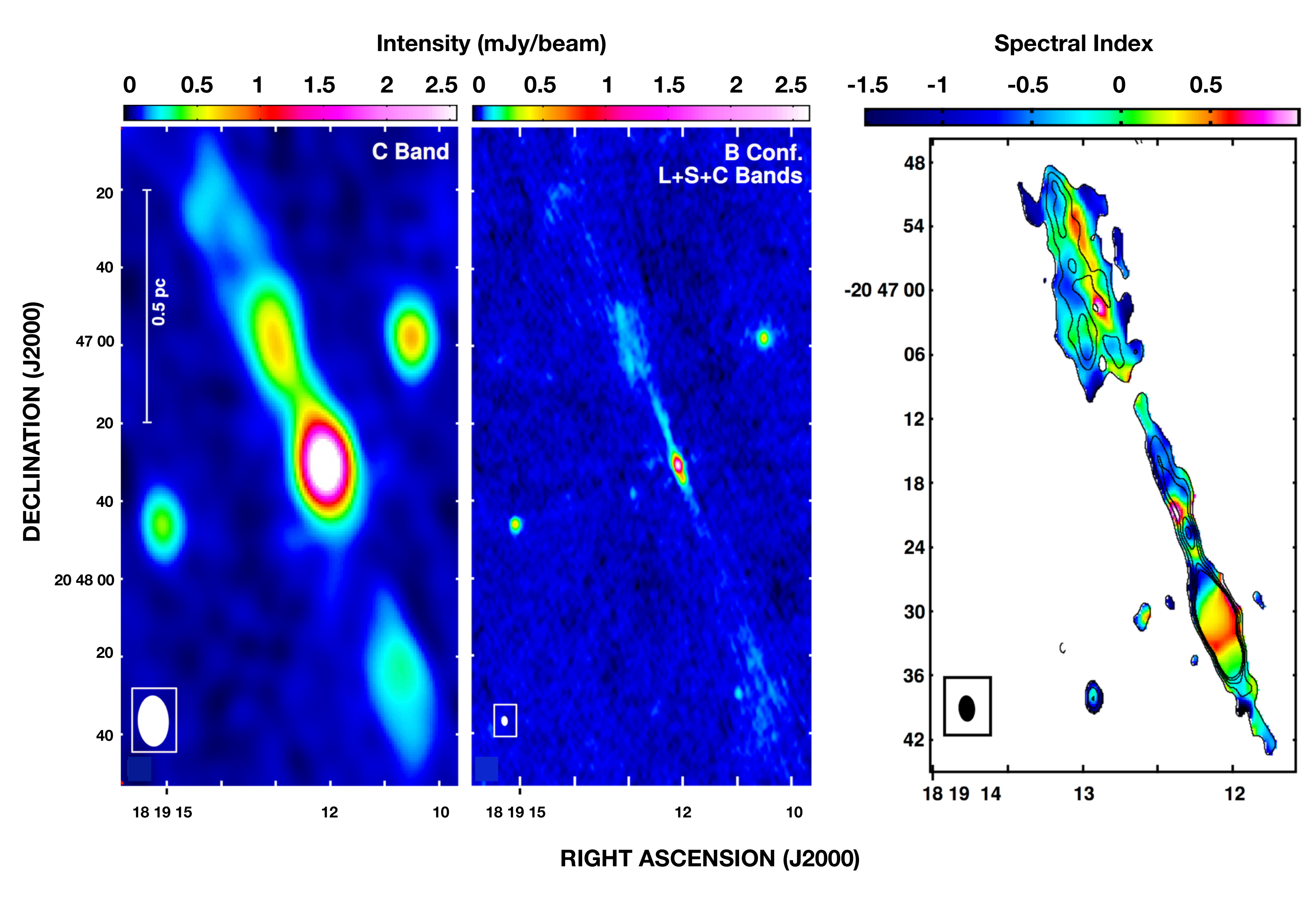}
\end{center}
\vspace{-1cm}
\caption{\footnotesize
The radio jet from IRAS 18162--2048, which powers Herbig-Haro objects 80 and 81. This object is the prototypical example of a radio jet from a high-mass YSO, well studied thanks to its proximity (1.7 kpc), brightness, and anomalously large (pc) extension. The {\it left} and {\it central} panels show low and high (arcsecond) angular resolution views of the thermal jet core and inner jet knots. The {\it right} panel shows the  spectral index derived from multifrequency synthesis imaging over the $L$, $S$ and $C$ bands. Positive values at the core and jet axis mark the thermal free-free component, while negative spectral indices pinpoint to non-thermal emission. Taken from \cite{RK17}. The ngVLA will allow us to perform this type of studies in statistically significant samples of radio jets from both nearby low-mass YSOs and high-mass YSOs at kpc distances. 
}
\label{fig:Jet_IRAS18162}
\end{figure*}

Thermal free-free emission from ionized gas is a ubiquitous signpost of  the physical processes associated with star formation. Such emission is complementary to and as important as that from dust and cold gas.
Massive (proto)stars can produce enough Lyman continuum photons to significantly ionize their own envelopes. The VLA was key  to the discovery and understanding of ultracompact (UC) and hypercompact (HC) HII regions: the small ($< 0.1$ pc $\approx 20 000$ au) pockets of ionized gas associated with newborn massive ($M_\star > 15~M_\odot$) stars \citep{DH81,HH81,Kurtz94,DePree14,Kalcheva18}. The nature of the fainter ionized sources detected by the VLA in massive star formation regions \citep{Rosero16,Moscadelli16} is not clear, however, since they are barely detected (fluxes $\lesssim 100~\mu$Jy at $\nu > 30$ GHz) and mostly unresolved (sizes $< 100$ mas, or $< 500$ au at 5 kpc).

\begin{figure*}
\begin{center}
\includegraphics[width=\textwidth]{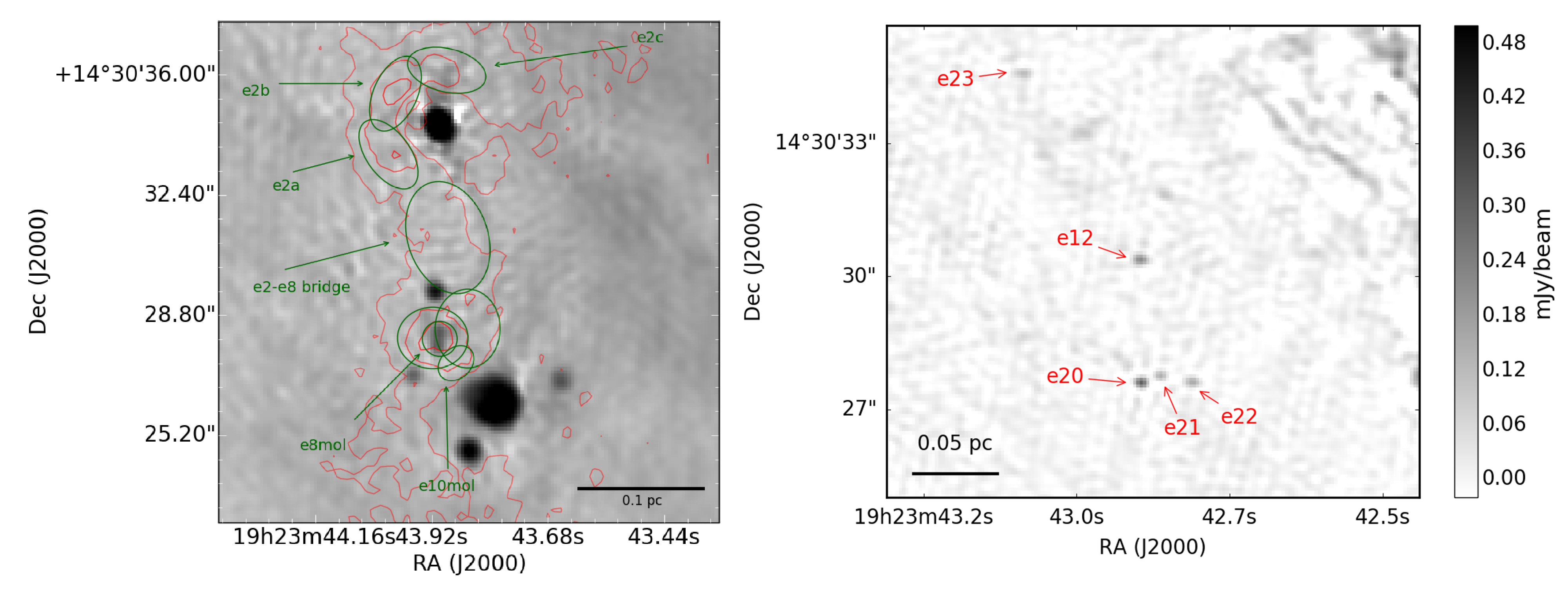}
\end{center}
\vspace{-1em}
\caption{\footnotesize
{\it Left:} Examples of bright hypercompact (HC) HII regions: the well-known e2/e8 system in the massive cluster forming region W51A. Gray-scale shows the free-free 2-cm continuum observed at 0.3 arcsec resolution. Contours show their surrounding molecular (formaldehyde) emission. Current facilities such as the VLA can map the continuum of all HII regions and the recombination line emission of some of these with $S_{\rm cm} \sim 10-100$ mJy. {\it Right:} Examples of faint ionized sources that are candidate HC HIIs, radio jets, or stellar-wind systems. Their fluxes $S_{\rm cm} \lesssim 0.1-1$ mJy and sizes $l < 100$ mas require the ngVLA to be characterized. Taken from \cite{Ginsburg16}. 
}
\label{fig:HIIs_W51}
\end{figure*}

Competing theoretical models suggest that these faint free-free sources could either be: i) gravitationally trapped HII regions \citep{Keto03}, expected to exist in the late stages of accretion in the formation of stars more massive than 20 to 30 $M_\odot$; ii) outflow-confined HII regions \citep{Tanaka16}, expected to form if there is a photoionized disk-wind outflow; or iii) radio jets similar to those found in their lower-mass counterparts \citep{Beltran16,Purser16}. 
Figure \ref{fig:Jet_IRAS18162} shows the iconic HH 80--81 radio jet, one of the brightest known, that is powered by the $\sim 10~M_\odot$ protostar IRAS 18162--2048 \citep{Girart17}. A significant fraction of the faint radio sources in massive star formation regions could be stellar winds from very young \citep{CG15} or unembedded massive stars \citep{Dzib13}, or even colliding-wind binaries with synchrotron excesses at $\nu < 4$ GHz \citep{Williams90,PM06,Ginsburg16}. Figure \ref{fig:HIIs_W51} illustrates the differences between the relatively bright HC HIIs in W51A, whose nature as photoionized regions around young massive stars is well established, and the compact, fainter sources that are candidates to be HC HIIs, thermal radio jets, or stellar-wind systems.

The high angular resolution and sensitivity of the ngVLA in both cm continuum and Radio Recombination Lines (RRLs) would open a new window to perform systematically detailed studies of the density and velocity distribution of the ionized gas present in the formation of massive stars. In Sections 2 and 3, we give more detailed sensitivity estimates for this scientific case.  Regardless of whether or not current theoretical models are correct in detail, free-free emission is the only way of tracing the stellar content in the deeply embedded regions where the formation of massive stars occurs.

{\it The Next Generation VLA will allow us to finally resolve the structure and kinematics of ionized material, the closest gaseous component to the (proto)star, in both low- and high-mass star formation regions.} 

\subsection{Free-free and Recombination Line Emission from Low-Mass Young Stellar Objects}
\begin{figure*}
\begin{center}
\includegraphics[width=0.8\textwidth]{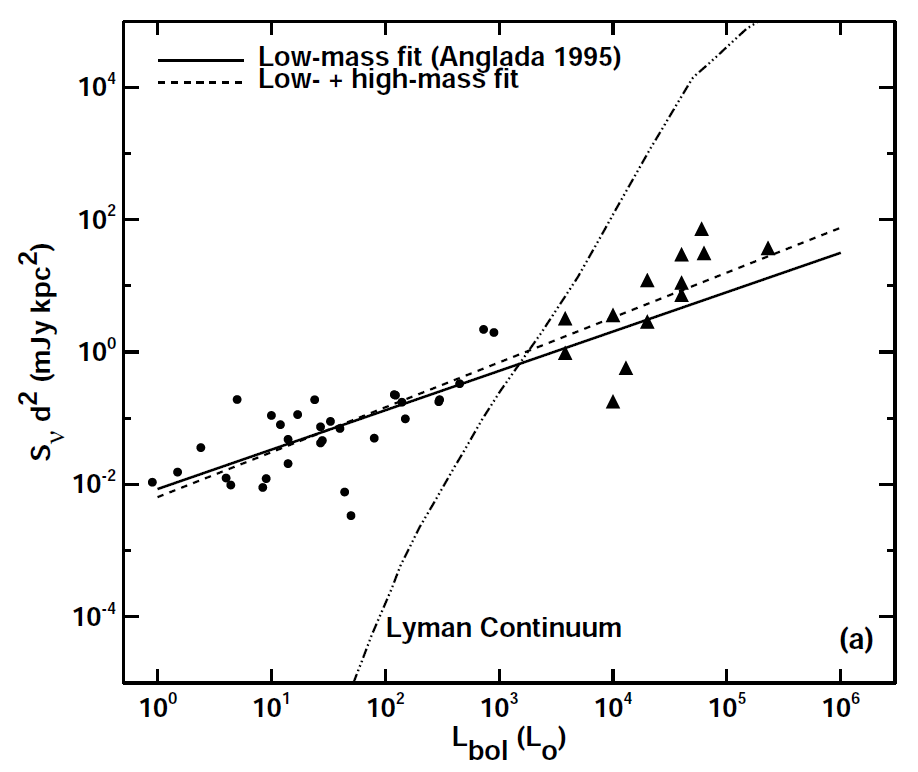}
\end{center}
\vspace{-1em}
\caption{\footnotesize
Empirical correlations between the YSO bolometric luminosity $L_{\rm bol}$ and the cm radio luminosity $Sd^2$. The plotted points and fits correspond to objects best interpreted as radio jets, from the low- to high-luminosity regime. The steeper dash-dotted line shows the expected radio luminosity for HII regions produced by a ZAMS star of the corresponding $L_{\rm bol}$. HC and UC HIIs are observed to populate this locus for  $L_{\rm bol} > 10^3~L_\odot$ (not plotted). The nature of radio objects close to the intersection -- of which exact value is somewhat model dependent -- is particularly difficult to determine (jet vs HII region). These objects at $\sim 10^3 - 10^4~L_\odot$ are highly relevant for understanding the formation of intermediate- and high-mass stars. The ngVLA will allow us to perform systematically resolved studies of radio jets in nearby star formation regions down to the brown-dwarf regime, and of HC HIIs at kpc distances down to $\lesssim 10^3~L_\odot$. Taken from \cite{CG10_PhD}. See also the recent review by \cite{Anglada18}. 
}
\label{fig:Correlations}
\end{figure*}

The VLA was the most important tool for the discovery of and understanding that low-mass Young Stellar Objects (YSOs) have at their centers collimated jets of free-free emission, known as thermal radio jets, powering  the observed molecular outflows and Herbig-Haro systems at larger scales \citep{Rodriguez94,Anglada95,Rodriguez97, Beltran01,Liu14,Tychoniec18}. This finding was an important confirmation of a key prediction of star formation theories \citep{Shu94}. Low-mass YSOs do not emit significant EUV photons from their photospheres. Thus, ionization produced by photons arising from the shock of the -- mostly -- neutral stellar wind against the surrounding high-density gas was established as a viable mechanism \citep{Torrelles85, Curiel87,Anglada98}. Figure \ref{fig:Correlations} shows a compilation of measurements of bolometric and radio luminosities of sources interpreted as being radio jets, compared to the theoretical expectations for their `jet vs HII region' nature. 

In the most embedded and youngest low-mass YSOs (Class 0 and I, or low-mass protostars), the radio jets are the only way of peering into the dense envelopes and tracing the ejection processes close (< 10 au) to the central objects. Even with the $\sim 10~\mu$Jy sensitivity and $\sim 10$ au resolution currently achievable with the VLA for the nearest low-mass star forming regions, however, our understanding of ejection phenomena in low-mas YSOs is far from complete. Specifically,  the relatively bright radio jets from Class 0 and I YSOs are barely resolved in the continuum, and their kinematics are unknown due to the practical impossibility of detecting their RRL emission. Furthermore, jet emission in the more evolved Class II YSOs -- the protoplanetary disks, now one of the most active fields in astronomy -- is far weaker and hardly detectable by the VLA in the most optimistic cases \citep{GalvanMadrid14, Macias16}. For Class II (primordial disks) and Class III (remnant disks) YSOs, other physical processes become observable. Disk photoevaporation, one of the main candidates to cause protoplanetary disk dispersal, may be detected through faint free-free and RRL emission \citep{Pascucci12,Owen13}. Also, unresolved non-thermal (gyro)synchrotron emission from active magnetospheres is often detected with the VLA in some of these more-evolved YSOs \citep{FW13,Liu14, Rivilla15}. The ngVLA will  enable us to resolve, for the first time, the structure and kinematics of the free-free emission from YSOs at all proto-stellar and young-stellar evolutionary stages from Class 0 to Class III, as well as disentangle it from other `contaminating' processes such as (gyro)synchrotron emission.  A full characterization of the free-free emission from YSOs at frequencies $\nu > 30$ GHz is also necessary to separate it from dust emission. This disentangling is a mandatory issue to address in studies of planet formation, which need to get accurate measurements of the mass distribution of disks around low-mass YSOs.

\section{Scientific Capabilities with Continuum Observations}

We estimate here the scientific outcome of deep (rms noise $0.5~\mu$Jy/beam) continuum total intensity observations from 1 to 50 GHz with an angular resolution of 30 mas or better. These values are about the ngVLA 1-hr sensitivity and resolution quoted in the project overview by \cite{Carilli15}. 
All the science cases described below require such wide frequency coverage to image the respective free-free emission and disentangle it from other `contaminating' emission mechanisms. At $\nu < 4$ GHz, there could be a contribution from (gyro)synchrotron emission that would need to be properly subtracted \citep[see chapter by][]{Hull18}. At $\nu > 50$ GHz dust emission starts to dominate in some objects. This transition region also needs to be covered to  disentangle properly the competing emission mechanisms \citep{Isella15}.  

A wide field-of-view and the ability to mosaic are needed to make studies over statistically significant samples, except for highly clustered regions \citep[e.g.,][]{Forbrich16}. A recent VLA survey of YSOs in a nearby star formation region needed 134 pointings to create a Nyquist-sampled mosaic of an area equivalent to 0.5 deg$^2$ (Hauyu Baobab Liu, personal communication). Under the assumption of 18-m dishes, the ngVLA would need about half the number of pointings than the current VLA. At the sensitivities considered in the following Sections, a full survey of the entire YSO population ($\sim 10^3$ sources) in a nearby star-forming cloud could be completed with a time investment of tens of hours.  

\subsection{Radio Jets from Low-Mass YSOs}

We use the empirical correlations between the radio luminosity of YSOs and their bolometric luminosity \citep{Anglada95,Shirley07,Anglada15}, which appear to hold even in the brown dwarf regime \citep{Morata15}. We also assume a distance of 140 pc throughout this section, characteristic of nearby low-mass star formation regions such as Taurus and Ophiuchus \citep{OrtizLeon17}. For the lowest-luminosity case, $L_{\rm bol}=0.01 ~ L_\odot$, this relation gives a 3.6 cm flux $F_{\rm 3.6cm} = 10~\mu$Jy, which is only detectable with deep VLA integrations. The jet is comprised of a jet core and extended emission typically in the form of jet knots (see Fig. \ref{fig:Jet_IRAS18162} for one of the few well studied cases). The jet core is usually the brightest and we consider it to have $\sim 50 \%$ of the total jet emission, ($F_{\rm 3.6cm, core} = 5~\mu$Jy). Such emission would be detectable easily with the ngVLA at S/N $\sim 10$ in $\sim 1$ hr integrations \citep{Carilli15}. 

Depending on its driving mechanism, evolutionary stage, and mass, the jet core emission can be spread through very different scales \citep{Frank14}, from magnetospheric (diameter $\sim 0.1$ au = 0.7 mas at 140 pc) in the X-wind model to disk-wind scales ($\sim 10$ au = 70 mas). The ngVLA with an angular resolution of 5-20 mas would be able to distinguish between jet launching models, detecting unresolved or barely resolved cores for the former scenario and resolved ones for the latter. It is also possible that some jets are collimated externally through ambient pressure or magnetic fields, in which case the emission would extend to tens of au \citep{Albertazzi14} and be more easily resolved. For an unresolved X-wind jet core, most of the flux would be inside the beam, so it would be easily detected. For an resolved disk-wind jet core, sensitivity to different parts of the emission is model dependent, but assuming a conservative electron temperature $T_e=5000$ K, then $T_B=50 - 5000$ K for $\tau=0.01 - 1$, so the fainter optically-thin emission could be detected at  S/N $\sim 50/7 = \sim 7$ for 1 hr integration at 10 GHz (Carilli et al. 2015).

\subsection{Photoevaporating Flows from Class II YSOs}

During the Class II stage, free-free emission from the radio jet becomes weaker due to the much lower accretion rates compared to the Class 0 and I stages. Free-free emission from disk photoevaporation by EUV photons could then take over as the principal continuum emission \citep{Pascucci14,GalvanMadrid14,Macias16}. The possibility of detecting and imaging this emission is highly relevant for understanding the mechanisms of protoplanetary disk dispersal \citep{Alexander14}. \cite{Pascucci12} applied the photoevaporation-flow model of \cite{Hollenbach94} to estimate the free-free continuum and hydrogen recombination line emission due to EUV photoevaporation. Using their prescription, disks with an ionizing photon luminosity as low as $\Phi=10^{40}$ s$^{-1}$ at 140 pc would have centimetric fluxes (almost constant with frequency) $S_{\rm cm} \sim 4~\mu$Jy. Therefore, EUV  photoevaporation happening in low-mass protoplanetary disks could be easily detectable with the ngVLA at S/N $\sim 10$ in 1-hour integrations \citep{Carilli15}.

\subsection{Faint Ionized Sources in High-Mass Star Formation}

The VLA has been the main tool for discovering and characterizing the small pockets of ionized gas around newly formed O-stars: the so-called ultracompact (UC) and hypercompact (HC) HII regions. In the following, we assume ionization equilibrium -- something debatable in the faintest sources that the ngVLA will help to understand --  and use the observational calibration for main-sequence OB stars of \cite{Vacca96}. We also assume a typical distance of 5 kpc for massive star formation regions throughout this section. A B0 main sequence star of $20~M_\odot$ has $L_{\rm bol}=10^{4.8}~L_\odot$ and Lyman continuum photon rate $N_{\rm Ly} = 10^{48.1}$ s$^{-1}$, which translates into a 8-GHz optically-thin flux  $\approx 480$ mJy (quite constant across frequencies since the dependency for optically thin emission is $\nu^{-0.1}$). 
This emission level indeed defined the target sensitivities for the past few decades \citep{WC89,Kurtz94,Urquhart09}.

Extrapolating the empirical calibration used above to lower masses, the HC HII of an intermediate-mass star with $L_{\rm bol}=10^3~L_\odot$ has an 8-GHz flux of $120~\mu$Jy \citep[e.g.,][]{Cesaroni15}. The VLA is currently allowing observers to detect, but not to resolve, a population of radio sources in massive star formation regions with centimeter fluxes $\lesssim 100~\mu$Jy \citep[e.g.,][]{Rosero16,Moscadelli16}. Their interpretation as faint HC HIIs, however, is not unique, and many of them can be as well interpreted as radio jets from high-mass YSOs. Extrapolating the radio-jet flux to bolometric luminosity correlation to high masses gives jet centimeter fluxes that are below HC HII fluxes for all bolometric luminosities above $L_{\rm bol} \sim 10^3~L_\odot$ (see Fig. \ref{fig:Correlations}, we note that the exact intersection is model dependent). Radio jets, however, are often assumed to have ionization fractions $\sim 0.1$ for low-mass YSOs, whereas UC and HC HIIs are considered to be fully ionized. The theoretically expected quenching of HC HIIs due to their surrounding neutral accretion flows \citep{Walmsley95,Peters10} and a higher ionization fraction in jets from intermediate- and high-mass YSOs could easily invert the relative radio brightnesses of the two different physical scenarios. Observationally, there is evidence that a jet nature may be the correct interpretation for the faint ionized sources of some YSOs with  luminosities up to $L_{\rm bol} \sim 10^4~L_\odot$ \citep{Rosero16,Moscadelli16,Purser16}. At these luminosities there could be a transition to HII-region-dominated emission \citep{Guzman14,Guzman16}.  

The ngVLA is needed to understand the nature of this population of  objects with cm fluxes $S_{\rm cm} \lesssim 100~\mu$Jy and sizes $l \lesssim 100$ mas (see Fig. \ref{fig:HIIs_W51}). Assuming a conservative electron temperature $T_e=5000$ K, then $T_B=50 - 5000$ K for $\tau=0.01 - 1$, so a 1 hour integration with an rms noise  $\sim 0.5~\mu$Jy/beam $\sim 10$ K would suffice to map very optically thin emission at $5 - 30$ mas resolution, which would correspond to $25 - 150$ au at 5 kpc.

\section{Scientific Capabilities with Recombination Line Observations}

We estimate here the scientific outcome of deep (rms noise of  $100~\mu$Jy/beam in channels 5 km/s wide) total intensity observations of RRLs from H$78\alpha$ (5.0 GHz) to H$40\alpha$ (99.0 GHz). The desired maximum angular resolution is 100 mas for unresolved detections, 30 mas for resolved detections, or slightly better for the few brightest cases. A hundred milliarcseconds correspond to 14 au for objects in nearby (140 pc) low-mass star forming regions, or to 500 au at 5 kpc. 

Continuum imaging is not enough to model and fully interpret ionized sources in star formation. The kinematic information  inferred from hydrogen recombination lines is also needed. Recombination lines in the cm radio and (sub)mm are weak compared to the often used optical and infrared lines such as H$\alpha$, Ly$\alpha$, and Br$\gamma$. They are unobscured, however, which makes RRLs a unique tool for revealing the kinematics in the embedded regions where star formation occurs \citep{MartinPintado89,Peters12,JimenezSerra13,Plambeck13,Zhang17}. 

The VLA, ALMA, and other interferometers have been able to detect systematically RRLs from the UCs and HC HIIs produced by O and early-B type stars \citep{KZK08}, but not for lower masses. The resolved kinematics of ionized gas in star formation is almost unexplored except for the brighter UC HIIs and some HC HIIs \citep[e.g.,][]{Garay86,Liu12}. Also, only a couple of detections of RRL emission from bright radio jets have been reported \citep{JimenezSerra11,JimenezSerra13}, probably with significant contribution from an underlying HII region. Attempts to detect RRLs in iconic objects such as the jet of IRAS 18162--2048 (Fig. \ref{fig:Jet_IRAS18162} shows its continuum emission) have produced negative results (Roberto Galv\'an-Madrid, unpublished). Indeed, RRLs are more easily detected in HC HIIs than in radio jets because the former tend to be brighter and their line widths are one order of magnitude smaller. Even `broad linewidth' HC HIIs have $\Delta V \sim 30 - 50$ km s$^{-1}$, whereas jet motions appear to have $\Delta V$ of several hundreds of km  s$^{-1}$, as inferred from their proper motions \citep{Curiel06,RK16,Guzman16} and a few tentative RRL detections \citep{JimenezSerra11}. Therefore, ngVLA measurements of  RRL line widths alone, even if angularly unresolved, would serve as a discriminator for competing models of the interpretation of faint ionized sources in massive star formation. 

 Lines such as H$51\alpha$ are detectable for $L_{\rm bol}=10^5~L_\odot$ objects even with the sensitivities of the VLA before its upgrade \citep[e.g.,][]{GalvanMadrid09}. In the following, we consider the detectability of RRL emission from fainter sources at kpc distances with the ngVLA. The LTE line-to-continuum ratio $S_{\rm line} \Delta V / S_{\rm cont}$ goes as $\nu^{1.1}$ \citep{GS02}. Observationally, it has been measured to be $S_{\rm line} \Delta V / S_{\rm cont} = 30 - 100$ km s$^{-1}$ for the H$30\alpha$ line (231.9 GHz) for HC HIIs with $L_{\rm bol} \sim 10^5 L_\odot$ \citep{KZK08}. We consider $S_{\rm line} \Delta V / S_{\rm cont} = 50$ km s$^{-1}$ for the H$30\alpha$ line and $\Delta V = 25$ km s$^{-1}$ for all lines. This estimate translates into $S_{\rm line} \Delta V / S_{\rm cont} \sim 17.6$ km s$^{-1}$ for H$41\alpha$ (90.0 GHz), 8.9 km s$^{-1}$ for H$51\alpha$ (48.1 GHz), and 0.8 km s$^{-1}$ for H$106\alpha$ (5.4 GHz). 
We further consider a HC HII region with a 48 GHz continuum flux $S_{\rm 48GHz}=1$ mJy, such as the ones detected in $L_{\rm bol} \sim 10^4~L_\odot$ YSOs by \cite{VdTM05}. This HC HII would have a velocity-integrated H$51\alpha$ flux $S_{\rm line} \Delta V=8.9$ mJy km s$^{-1}$. At this frequency, the expected ngVLA noise in a 5 km s$^{-1}$ channel for a 10-hour integration is $\sim 32~\mu$Jy/beam  \citep{Carilli15}. For the assumed FWHM linewidth of 25 km s$^{-1}$, the line peak would be $S_{\rm peak} \sim 334~\mu$Jy, detectable at S/N $\sim 10$. 

The above discussion applies to beam-matched or angularly unresolved observations with a HPBW $\sim 100$ mas. Observations at higher angular resolution would need to aim at brighter objects or integrate longer. 
The S/N of individual line detections for these faint HII regions would go up by factors $\sim 2$ for the H$41\alpha$ line at 90 GHz, so resolved RRL mapping would be less expensive using high-frequency RRLs. Lines in the frequency range $50 < \nu < 100$ GHz could be the right compromise between high brightness and the absence of contamination from molecular lines compared to RRLs at ALMA frequencies. Therefore, RRL observations are an important motivation for the ngVLA to operate at frequencies $\nu > 50$ GHz. 

Non-detections at frequencies $\nu < 20$ GHz or in fainter objects could be solved by taking advantange of stacking of lines from neighbouring quantum numbers. The VLA Ka and Q bands ($36 - 50$ GHz) harbor 12 $\alpha$ hydrogen RRLs (H$62\alpha$ to H$51\alpha$) that could be stacked together in a given object to decrease the noise by factors of $\sim 3 - 4$. This opportunity highlights the importance of having a large simultaneous frequency coverage. 

Although the previous discussion considered a HC HII from a massive (proto)star at kpc distances, it also applies to radio jets from low-mass YSOs at the canonical distance of 140 pc. These radio jets have cm fluxes $S_{\rm cm} \sim 1$ mJy for  luminosities $L_{\rm bol} > 10~L_\odot$ (see Fig. \ref{fig:Correlations}). Their RRL emission could be spread through several hundred km s$^{-1}$. Assuming that the radio-jet line width is $\times10$ larger than for HC HIIs, the S/N would decrease by factors of $\sim 10^{0.5} \sim 3$ if the velocity binning increases from $5 - 50$ km s$^{-1}$. 

In conclusion, although more expensive than continuum observations, kinematical imaging of hydrogen recombination line emission is feasible with the ngVLA. Such observations would open for the first time a window into understanding the kinematics of ionized gas in the formation of both low- and high-mass stars. 

\section{Peering into the Innermost Regions in the Formation of Stars of all Masses}

From the previous sections, it can be concluded that the ngVLA will allow us to perform, for the first time, studies of statistically significant samples of the resolved structure and kinematics of the ionized gas associated with  stars in formation and young stars of all masses. This ionized gas is often the closest gaseous component to the (proto)star itself. Table \ref{table:1} summarizes the basic continuum properties of the objects that we have considered in detail among the anticipated `radio zoo'.

\begin{table*} 
\centering
{
\renewcommand{\arraystretch}{1.1}
\begin{tabular}{@{} l *4c  @{} }
Object & Typical distance & 8 GHz Flux & Size  \\
	   &  [kpc]  &  [$\mu$Jy]   & [mas (au)]  \\
\hline	   
Faint HC HII  &    &    &        \\
or massive radio jet  &  5   &  $< 100$  & $< 100$ ($< 500$)   \\
\hline
Very low  &  & & \\ 
luminosity radio jet & 0.14  &  5  & 0.7 -- 70 (0.1 -- 10) \\ 
\hline
Photoevaporation flow &  &  & \\
from low-mass disk  &  0.14 & 4 & 7 -- 350 (1 -- 50) \\
\hline
\end{tabular}
\caption{Free-free continuum properties of the considered objects.} \label{table:1}
 }
\end{table*}

\section{Uniqueness of ngVLA Capabilities}

Only the VLA and a future ngVLA can observe the free-free continuum emission associated with star formation in the range of frequencies where it dominates ($4 - 50$ GHz). At frequencies $< 4$ GHz and $> 50$ GHz, (gyro)synchrotron and dust emission, respectively, can be a significant `contaminant', or even dominate the observed flux. Having a telescope with frequency coverage beyond the overlapping ranges is needed to model properly and subtract the contribution to the emission from those other mechanisms. Conversely, properly subtracting the free-free contribution will be mandatory in the much anticipated studies of dust emission in protoplanetary disks. 

Similarly, hydrogen radio recombination lines  from about 20 to 100 GHz have the right compromise between brightness (they are too faint at lower frequencies) and the absence of confusion with the rich molecular forest seen at higher (ALMA) frequencies. 

A field-of-view of at least a few arcminutes size, as given by the proposed 18-m dishes, is needed to perform surveys of large numbers of sources in areas of the order of a square degree in reasonably long  integrations of tens of hours. 

Although not discussed at length in this paper  \citep[see chapter by][]{Hull18}, the capability to image in full polarization mode is highly desirable, since a few radio jets are known to have synchrotron emission beyond the thermal core \citep{CG10}, and that might be the rule rather than the exception.  

Finally, a large instantaneous bandwidth (of the order of the central frequency where possible) would uniquely open the time-domain window in a systematic way. Radio jets \citep{Liu14} and some HC HIIs \citep{DePree14} are known to be variable in timescales from days to years, a phenomenon with important implications to the properties of accretion in YSOs \citep{GM11,Hunter17}.    

\section{Synergies at other Wavelengths}

The ngVLA would naturally complement and enhance the star formation research done with ALMA and the SKA. The studies of free-free and recombination line emission proposed in this chapter will consistently be put in the context of ALMA observations of the corresponding dust and molecular line emission. All tracers are needed to have a comprehensive physical picture. SKA observations at lower-frequencies would greatly help to disentangle non-thermal emission mechanisms. The proposed ngVLA research would also benefit observations with the next generation of ground- and space-based near- and mid-IR telescopes, such as the TMT, GMT, E-ELT, JWST, and WFIRST, since those  observations are the most efficient to discover and classify YSOs. Finally, high-angular resolution mid-IR and far-IR observations, such as the ones provided by SOFIA and other future facilities, are in high demand to pinpoint the sources of luminosity at resolutions that are not too coarse compared with instruments like ngVLA, ALMA, and JWST.

\acknowledgements 
The National Radio Astronomy Observatory is a facility of the National Science Foundation operated under cooperative agreement by Associated Universities, Inc. RGM acknowledges support from UNAM-PAPIIT program IA102817.



\end{document}